\documentclass[a4paper]{article}
\usepackage{amssymb, bbm, dsfont}
\usepackage{pifont, url}
\newcommand{\cmark}{\ding{51}}%
\newcommand{\xmark}{\ding{55}}%

\usepackage{INTERSPEECH2021}

\title{Streaming Multi-talker Speech Recognition with Joint Speaker Identification}
\name{Liang Lu, Naoyuki Kanda, Jinyu Li and Yifan Gong}
\address{Microsoft Corp., USA}
\email{\{liang.lu, naoyuki.kanda, jinyli, yifan.gong\}@microsoft.com}

\begin{document}

\maketitle
\begin{abstract}
In multi-talker scenarios such as meetings and conversations, speech processing systems are usually required to transcribe the audio as well as identify the speakers for downstream applications. Since overlapped speech is common in this case, conventional approaches usually address this problem in a cascaded fashion that involves speech separation, speech recognition and speaker identification that are trained independently. In this paper, we propose Streaming Unmixing, Recognition and Identification Transducer (SURIT) -- a new framework that deals with this problem in an end-to-end streaming fashion. SURIT employs the recurrent neural network transducer (RNN-T) as the backbone for both speech recognition and speaker identification. We validate our idea on the LibrispeechMix dataset -- a multi-talker dataset derived from Librispeech, and present encouraging results. 
\end{abstract}

\section{Introduction}

In~\cite{morgan2003meetings}, Morgan {\it et al.} referred to the process of spoken language from meetings as a nearly ``ASR-complete" problem. The authors used this metaphor to highlight that almost all the key problems in spoken language process can be explored in the context of meetings, such as conversational speech recognition, signal processing with far-field microphone arrays, overlapped speech detection and separation, utterance segmentation and speaker identification or diarization,  disfluency detection, etc. Indeed, rich transcription of meetings and conversational speech is still a challenging problem to the speech community nowadays. Since the birth of the ICSI meeting corpus~\cite{janin2003icsi} and AMI dataset~\cite{renals2007recognition}, the progress to deal with this challenge has been made mainly by following the divide-and-conquer strategy, namely, researchers usually attack each sub-problem independently, and put all the components together in a cascaded fashion for meeting transcription~\cite{stolcke2007sri, hain2011transcribing}. Not surprisingly, such systems are difficult to be optimized end-to-end, and their computational complexities are usually very high. More importantly, such systems are usually not capable to be deployed in a streaming scenario with a low latency requirement. 

Since deep learning reshaped the landscape of the speech field~\cite{hinton2012deep}, recent years have witnessed the paradigm shift for almost every problem covered by the meeting transcription scenario, such as automatic speech recognition (ASR)~\cite{graves2013speech, chorowski2015attention, chan2016listen, lu2016training, he2019streaming}, speaker verification~\cite{heigold2016end, snyder2016deep}, speech separation~\cite{hershey2016deep, yu2017permutation}, and speaker diarization~\cite{garcia2017speaker, zhang2019fully, fujita2019end} etc. As a result, the systems for meeting transcription have evolved remarkably~\cite{yoshioka2018recognizing}. However, these systems mostly still operates in a cascaded fashion, and have inherited the drawbacks of their previous generation. Recently, an emerging trend in the speech community is to address the problems in meetings in an end-to-end fashion, and jointly optimize all the components. In particular, there are series of research studies on end-to-end multi-talker ASR that deal with speech separation and speech recognition jointly~\cite{settle2018end, chang2019end, kanda2020serialized, tripathi2020end}. While those works are mostly constrained in an offline condition, in~\cite{lu2020streaming, sklyar2020streaming}, the authors investigated steaming multi-talker ASR based on recurrent neural network transducers (RNN-T), and reported promising results. 

In this paper, we aim at a further stride along this direction, and explore joint speech separation, ASR, and speaker identification (SID) with a streaming, end-to-end model. Built upon our prior work on streaming multi-talker speech recognition~\cite{lu2020streaming}, we propose a novel approach referred as Streaming Unmixing, Recognition and Identification Transducer (SURIT) for this problem. While streaming end-to-end speech recognition has achieved a remarkable progress, most works on speaker identification only concern offline scenarios. In order to perform speech recognition and speaker identification jointly in a streaming fashion, we introduce RNN-T based speaker identification. Ultimately, similar to~\cite{lu2020streaming}, SURIT firstly applies an Unmixing module to extract single-speaker feature representations, followed by RNN-Ts for streaming transcription and speaker identification. To evaluate our proposed model, we performed experiments using the LibriSpeechMix dataset~\cite{kanda2020serialized}, which simulate the overlapped speech data from the LibriSpeech corpus~\cite{panayotov2015librispeech} with an artificially defined speaker identification protocol. We show that SURIT can achieve strong results for both speech recognition and speaker identification.


\section{Related Work}


There were limited studies on the joint modeling of end-to-end ASR and SID. One simple approach is inserting a speaker-role tag \cite{el2019joint} or a speaker-identity tag \cite{mao2020speech} at the end of each utterance. However, \cite{el2019joint} and \cite{mao2020speech} did not cope with overlapping speech and also their approach cannot identify a speaker who is not included in the training data. Recently, Kanda et al.~\cite{kanda2020joint} proposed an offline sequence-to-sequence (S2S) based approach that can transcribe and identify the speakers from overlapping speech given a set of speaker profiles. 
Our work differs from~\cite{kanda2020joint} mainly in that SURIT applies the RNN-T for both ASR and SID, which can run in a streaming fashion. The idea of using an RNN-T for SID is inspired by~\cite{waters2019leveraging}, which proposed RNN-T based language identification. Our speaker verification module depends on a speaker inventory as an auxiliary information as same with \cite{kanda2020joint}.
Overall, 
our work is the first study on RNN-T based SID that is trained jointly with multi-talker ASR in a {\it streaming} fashion. 

\section{RNN-T: Review}
\label{sec:rnnt}

RNN-T is the backbone in our model, which directly computes the sequence-level conditional probability of the label sequence given the acoustic features by marginalizing all the possible alignments. Given an acoustic feature sequence $X=\{x_1, \cdots, x_T\}$ and its corresponding label sequence $Y=\{y_1, \cdots, y_U\}$, where $T$ is the length of the acoustic sequence, and $U$ is the length of the label sequence, RNN-T defines the conditional probability 
\begin{align}
\label{eq:prob}
P(Y \mid X) = \sum_{\tilde{Y} \in \mathcal{B}^{-1}(Y)}{P(\tilde{Y} \mid X)},
\end{align}
where $\tilde{Y}$ is a path that contains the blank token $\O$, and the function $\mathcal{B}$ denotes mapping the path to $Y$ by removing the blank tokens in $\tilde{Y}$. The probability can be efficiently computed by the forward-backward algorithm, which requires to compute the probability of each time step. Let $f(\cdot)$ denote the audio encoder, and $g(\cdot)$ the label encoder, the feature representations can be obtained as
\begin{align*}
\bm{h}_{1:t}^f = f(x_{1:t}), \hskip4mm \bm{h}_{1:u}^g = g(y_{0:u-1}),
\end{align*}
where $y_0$ is a blank token. RNN-T fuses the two feature representations by a joint network $j(\cdot)$ as
\begin{align}
\label{eq:joint}
\bm{z}_{t, u} = j({\bm h}_t^f, \bm{h}_u^g),
\end{align}
and the conditional probability of the next token is obtained by a Softmax function, i.e., 
\begin{align}
P(y_u \mid \bm{z}_{t, u}) = \text{Softmax}(\bm{W}\bm{z}_{t,u} + \bm{b}),
\end{align}
where $\bm{W}$ and $\bm{b}$ are the weight matrix and bias. Given $P(y_u \mid \bm{z}_{t, u})$, the sequence-level conditional probability can be computed by dynamic programming, and the RNN-T loss can be defined as the negative log-likelihood as:
\begin{align}
\label{eq:nll}
\mathcal{L_{\text{rnnt}}} (Y, X) = -\log P(Y \mid X)
\end{align}

\begin{figure}[t]
\small
\centerline{\includegraphics[width=0.3\textwidth]{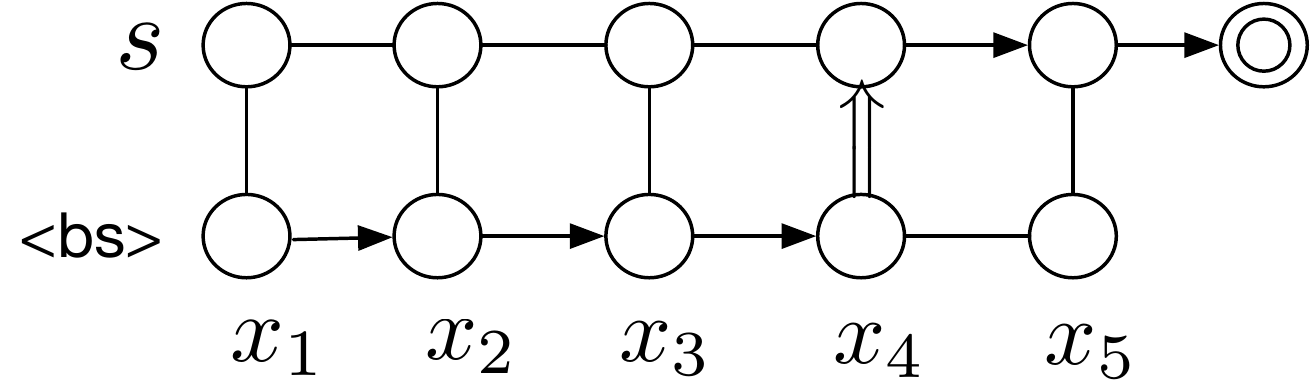}}
\caption{An illustration of using RNN-T for streaming speaker identification. The arrowed path corresponds to \{$\langle bs\rangle \langle bs\rangle \langle bs\rangle s \langle bs\rangle$\}, where $\langle bs\rangle$ denotes the blank speaker token. The probability of the arc $\rightarrow$ is $b_{t,u}$, and the probability of the arc $\Rightarrow$ is $(1-b_{t,u})P(s \mid \bm{z}_{t,u})$ in Eq.~\eqref{eq:prob2}.}  
\label{fig:sid}
\vskip -5mm
\end{figure}

\section{Streaming SID with Transducers}

In this paper, we propose to apply transducers for streaming SID. 
Here, we assume a single-talker audio to
introduce the idea of transducer-based SID.
It will be extended for the joint modeling with multi-talker ASR with speech overlap in Section \ref{sec:joint_ASR_SID}.

Suppose we have a speaker inventory $D=\{d_k|k=1,...,K\}$, where $K$ is the total number of the profiles, $d_k$ is a speaker embedding (e.g., d-vector \cite{variani2014deep}) of the $k$-th speaker in the speaker inventory. Given an input acoustic feature $X$, the goal of SID is estimating the speaker index $k$ corresponding to $X$.
To simulate the real scenario, the speaker inventory $D$ can be {\it different} for each audio segment $X$, indicating that the participants of the meeting or conversation may vary each time. Standard SID approaches typically addresses this problem by extracting the vector representation of $X$, and measure its distance against the $d$-vectors in $D$ according to a metric such as the cosine distance. However, those approaches usually operate in an offline fashion and require a voice activity detector (VAD) if the audio is not segmented. On the other hand, in this work, we focus on streaming SID that can emit the speaker label before the utterance is ended, and can handle both silent and voiced regions without a VAD. 

\subsection{Transducer-based SID}

For the proposed transducer-based SID, we treat each speaker index $s\in \{1,...,K\}$ as the token that should be predicted by the RNN-T, 
as illustrated in Fig.~\ref{fig:sid}. 
To tailor the RNN-T model to fit in this scenario, we need to accommodate a few model structure changes. Since we do not capture the dependencies among the speaker labels, 
we omit the recurrent layers in the label encoder of RNN-T. To train such a model, we need to calculate the probabilities of emitting both the speaker label $s$ and the blank token $\langle bs \rangle$.
Given $\bm{z}_{t,u}$ from Eq.\eqref{eq:joint}, the probability of emitting the speaker label $s$ can be represented as
\begin{align}
\label{eq:spk-pro}
P(s = k \mid \bm{z}_{t,u}) = \frac{\exp (d_k^\top\bm{z}_{t,u})}{\sum_{k^\prime=1}^K{\exp (d_{k^\prime}^\top \bm{z}_{t,u})}},
\end{align}
Where $d_k$ is the d-vector of the target speaker.
However, in the speaker inventory, there is no embedding vector for $\langle bs \rangle$, so we cannot apply Eq.~\eqref{eq:spk-pro} to compute the probability of $P( \langle bs \rangle \mid \bm{z}_{t,u})$. To deal with this problem, we apply a special type of RNN-T model, i.e., Hybrid Autoregressive Transducer (HAT)~\cite{variani2020hybrid}, which was originally proposed for external language model integration for RNN-T. The key idea of HAT is that it reformulates the token probability, and uses two distributions for blank and label tokens, i.e.,
\begin{align}
\label{eq:prob2}
P(v_u \mid \bm{z}_{t,u}) =\left\{
 \begin{array}{ll}
 b_{t,u} & v_u = \langle bs \rangle \\
 (1-b_{t,u})P(s \mid \bm{z}_{t,u})  & \text{otherwise}
 \end{array}
 \right.
\end{align}
where $b_{t,u}$ is a Bernoulli distribution that can be estimated using a Sigmoid function, while  $P(s \mid \bm{z}_{t,u})$ is defined as Eq.~\eqref{eq:spk-pro}. Note that the value space of $s$ is $[1, K]$, while the value space of $v_u$ is $[1, K] \cup \langle bs \rangle$. With HAT, we can compute the emission probability of $\langle bs \rangle$ without its d-vector. 

\subsection{Optimizing the SID latency}
\label{sec:latency}

Naively applying the RNN-T loss as Eq.~\eqref{eq:nll} could cause the high latency of the HAT-based SID model.
The reason is that the vanilla RNN-T loss does not encourage the model to emit the labels early. More critically, for each path in Fig.~\ref{fig:sid}, the number of blank tokens is $T-1$ while there is only one speaker label for an audio input of $T$ frames. Consequently, the gradient of $\langle bs \rangle$ is significantly larger than the gradient of speaker labels. This encourages the model to prefer emitting $\langle bs \rangle$ during inference. To counteract this effect, we scale the gradient of $\langle bs \rangle$ as
\begin{align}
    \frac{\partial\tilde{ \mathcal{L}}}{\partial b_{t,u}} = \alpha \frac{\partial \mathcal{L}}{\partial b_{t,u}}, \hskip4mm \alpha \in (0, 1],
\end{align}
which is very similar to FastEmit~\cite{yu2020fastemit}, but we scale the gradient of $\langle bs \rangle$ since it is dominant. Inspired by the method to optimize the latency of end-pointer~\cite{li2020towards}, another approach is to penalize the probability of emitting the speaker label late, i.e.,
\begin{align}
\label{eq:latency}
    \log P(s \mid z_{t, u}) -= \max(0, \beta(t - t_{\text{buffer}} - t_{\text{delay}})),
\end{align}
where $\beta$ is a scaling parameter, $t_{\text{buffer}}$ is the grace period, and $t_{\text{delay}}$ means the delay when we simulate the overlapped audio. This equation means that for a path in Fig.~\ref{fig:sid}, if the time step $t$ that emits the speaker label is larger than $(t_{\text{buffer}} + t_{\text{delay}})$, then the probability of this path will be penalized. 

\begin{figure}[t]
\small
\centerline{\includegraphics[width=0.4\textwidth]{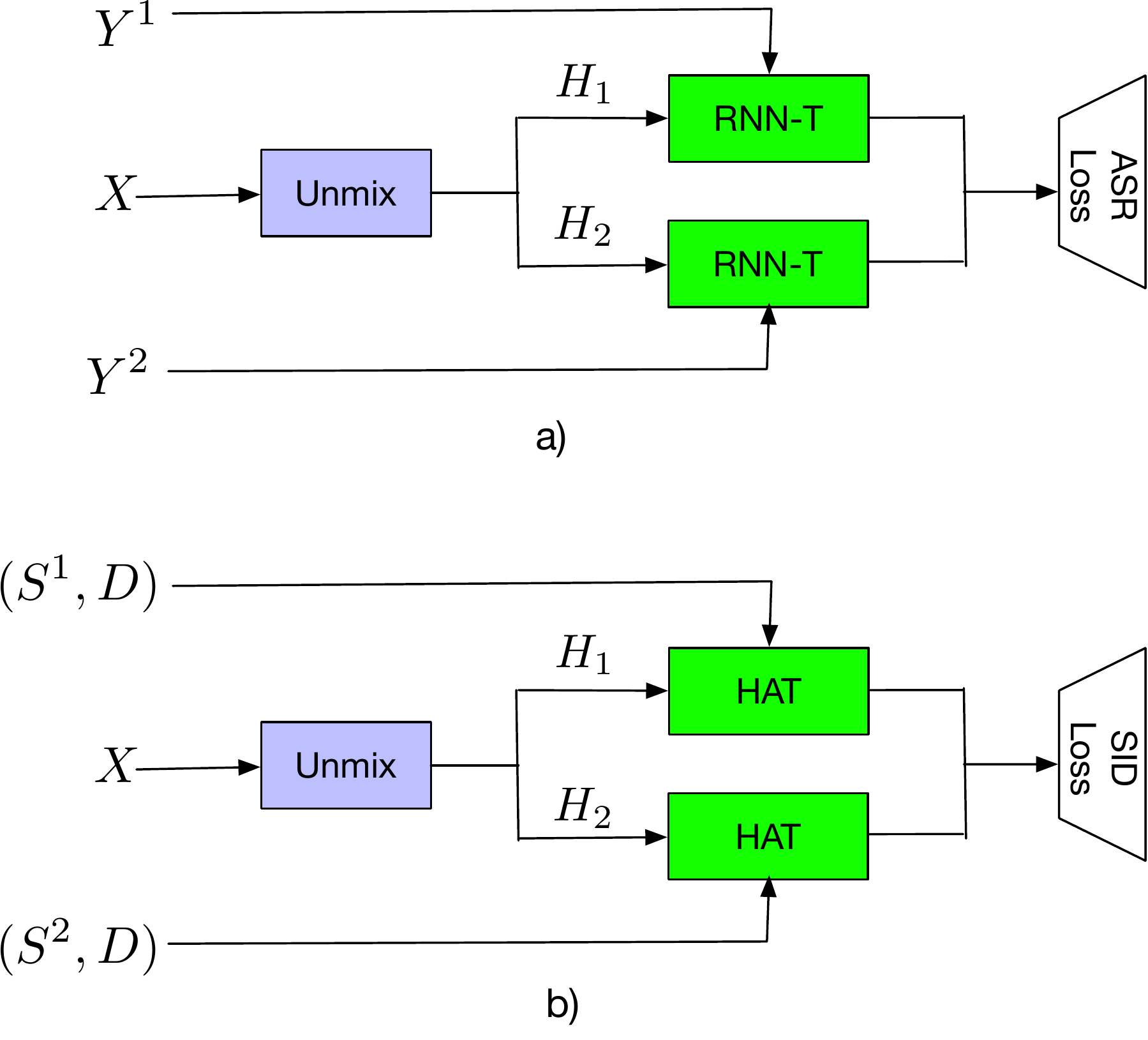}}
\vskip-4mm
\caption{a) The multi-talker speech recognition network. $Y^1$ and $Y^2$ are wordpiece labels, and $X$ represents acoustic features. b) The multi-talker speaker identification network. $S^1$ and $S^2$ are speaker labels, and $D$ refers to the speaker inventory. Note that, there are only one set of RNN-T and HAT model parameters, which are used to process the two feature streams from the Unmix module in the 2-speaker case. The Unmix module is shared by the speech recognition network and the speaker identification network. }  
\label{fig:surt}
\vskip -6mm
\end{figure}

\section{Joint Multi-talker ASR and SID}
\label{sec:joint_ASR_SID}

In a multi-talker scenario, dealing with the overlapped speech is the key challenge. In~\cite{lu2020streaming}, we proposed a streaming end-to-end model for multi-talker speech recognition, referred as SURT. The idea of SURT is illustrated in Figure~\ref{fig:surt}-a) for the 2-speaker case, in which, the Unmixing module emits two streams of audio representation where each stream contains the feature representations of one speaker. In this work, we use convolutional neural networks (CNNs) for the Unmixing module, in particular
\begin{align}
 \begin{array}{ll}
 M =  \sigma(\text{CNN}_\text{mask}(X)),  & H = \text{CNN}_\text{enc}(X), \\
H_1 = H * M,  & H_2 = H - H_1,
 \end{array}
\end{align}
where $\sigma$ denotes the Sigmoid function; $\text{CNN}_\text{mask}$ and $\text{CNN}_\text{enc}$ are two CNN encoders. The two feature representations are then fed into the same RNN-T network to compute the corresponding losses. It is well understood that label permutation is a key challenge for multi-talker speech processing~\cite{yu2017permutation}. Suppose the two ASR label sequences are $Y^1$ and $Y^2$, it is unclear if $H_1$ corresponds to $Y^1$ or $Y^2$, and similar for $H_2$. A widely used approach to address this problem is Permutation Invariant Training (PIT)~\cite{yu2017permutation}, which considers all the possible label assignments. In~\cite{lu2020streaming}, we evaluated another simple label assignment strategy, referred as Heuristic Error Assignment Training (HEAT)~\cite{tripathi2020end}. Unlike PIT, HEAT only considers one possible label assignment based on some heuristic information, for example, the order that they were spoken according to their starting times. Suppose $Y^1$ is from the speaker who spoken first, HEAT defines the loss function as
\begin{align}
\mathcal{L}_{\text{asr}}(X, Y^1, Y^2) = \mathcal{L}_{\text{rnnt}}(Y^1, H_1) + \mathcal{L}_{\text{rnnt}}(Y^2, H_2),
\end{align}
which means that we always assign the feature representation $H_1$ to the first spoken speaker. The loss function can drive the network to learn the corresponding feature mapping. In~\cite{lu2020streaming}, we show that HEAT works slightly better than PIT in terms of recognition accuracy, and it consumes less memory with lower computational cost. 

Similarly to SURT, 
we can represent the SID network for multi-talker case as shown in Fig.~\ref{fig:surt}-b). 
We can also define the speaker identification loss with HEAT as
\begin{align}
\mathcal{L}_{\text{sid}}(X, S^1, S^2) = \mathcal{L}_{\text{hat}}(S^1, H_1) + \mathcal{L}_{\text{hat}}(S^2, H_2),
\end{align}
where $S^1$ is the speaker label of the first spoken speaker, and $\mathcal{L}_{\text{hat}}(S^1, H_1)$ refers to the loss function defined by the HAT network. Given the two loss functions, we can train the two networks jointly by interpolating the two loss functions as
\begin{align}
\label{eq:joint_tr}
\mathcal{L} = \mathcal{L}_{\text{asr}}(X, Y^1, Y^2) + \lambda \mathcal{L}_{\text{sid}}(X, S^1, S^2),
\end{align}
where $\lambda$ is the interpolation weight.
Note that the Unmixing module is shared both for the ASR and SID networks.

\section{Experiments and Results}
\label{sec:exp}

\subsection{Dataset}
\label{sec:data}

Our experiments were performed on the LibriSpeechMix\footnote{\scriptsize \url{https://github.com/NaoyukiKanda/LibriSpeechMix}} dataset~\cite{kanda2020serialized}, which is the simulated overlapped audio derived from the LibriSpeech corpus~\cite{panayotov2015librispeech}. We used the same protocol to simulate the training and evaluation data as in~\cite{kanda2020serialized}, and we focus on the 2-speaker case. To generate the simulated training data, for each utterance in the original LibriSpeech {\tt train\_960} set, we randomly picked another utterance from a {\it different} speaker, and mix the latter with the previous one with a random delay sampled from 0.5 seconds and the length of the first utterance. For each overlapped audio segments, the size of the speaker inventory varies within the range $K \in [2, 8]$ for training including the 2 speakers presented in the overlapped audio, while for evaluation, we fixed $K=8$ and the minimum delay is 0 when mixing the two utterances. The dimension of the d-vectors in our experiments is 128. To generate the d-vectors in $D$, we randomly select 10 utterances for each speaker in Librispeech, and fed them into a pre-trained 17-layer convolutional network trained on VoxCelab~\cite{nagrani2017voxceleb, chung2018voxceleb2}. The model structure of the d-vector extractor is the modified version of the one in~\cite{zhou2019cnn}. We used the same approach to generate the evaluation dataset.


\subsection{Experimental Setup}
\label{sec:setup}

In our experiments, we used the 80-dimensional log-mel fiterbanks (FBANKs) as features, which were sampled the features at the 10 millisecond frame rate. We then spliced the features by a context window of 3 and finally downsampled super-frames by a factor of 3. Before feeding the features into the CNN blocks in the Unmixing module, we reshaped the input tensor to have 3 input channels, and consequently, the feature dimension became 80. For tokenization, we used 4,000 word-pieces as output tokens for RNN-T, which are generated by byte-pair encoding (BPE)~\cite{sennrich2015neural}. We used a 4-layer 2-dimensional CNN network for both encoder and mask component in the Unmixing module. We used a 6-layer LSTM as the audio encoder and 2-layer LSTM as the label encoder for the ASR RNN-T network, and a 1-layer LSTM for the audio encoder in the SID HAT network. The number of hidden units was set to be 1024 for all LSTM layers unless specified otherwise. The number of model parameters without the SID network is around 81 million, while the number of parameters in the SID network is around 9 million. We set the dropout ratio as 0.2 for LSTM layers, and applied one layer of time-reduction to reduce the input sequence length by the factor of 2~\cite{graves2012hierarchical, chan2016listen, lu2016segmental}. We also applied speed perturbation for data augmentation by creating two additional versions of the acoustic features with the speed ratios as 0.9 and 1.1. All our experiments are targeting the 2-speaker scenario. 
 
 We report the word error rate (WER) results by considering all the possible label permutations as in~\cite{tripathi2020end, kanda2020serialized, lu2020streaming, sklyar2020streaming}. To measure the SID accuracy, we follow the similar protocol as we calculate the WERs. Suppose the SID outputs of the two feature streams $(H_1, H_2) $ are $(Q^1, Q^2)$, we compute the speaker error rate (SER) as follows
\begin{align}
\text{SER} = \frac{1}{2N}\sum_{n=1}^N\min\Big(& \mathcal{E}\left([Q_n^1, Q_n^2], [S_n^1, S_n^2]\right), \nonumber \\
& \mathcal{E}\left([Q_n^2, Q_n^1], [S_n^1, S_n^2]\right)\Big),
\end{align}
where $n$ is the utterance index; $N$ is the total number of utterances; $\mathcal{E}(\cdot)$ refers to the edit distance function, and $[S_n^1, S_n^2]$  are the reference speaker labels of the $n-$th utterance. 

\subsection{Multi-talker ASR Results}
\label{sec:reco}

\begin{table}[t]\centering
\caption{Multi-talker speech recognition results. }
\label{tab:asr}
\footnotesize
\vskip-2mm
\begin{tabular}{cccc|c}
\hline 

\hline
    System    & Size & Feature    & Streaming  & WER \\ \hline
  PIT-S2S~\cite{kanda2020joint}    & 160.7M & FBANK &   \xmark   & 10.3   \\
  SOT-S2S-MBR~\cite{kanda2020minimum} & 135.6M & FBANK & \xmark & 8.3$^\dagger$ \\
  SURT~\cite{lu2020streaming}    & 81M & STFT     & \cmark  & 10.8   \\
  PIT-MS-RNN-T~\cite{sklyar2020streaming} & 80.9M & FBANK     & \cmark  & 10.2   \\
  SURT (this work) & 81M & FBANK     & \cmark  & 10.3   \\
  SURT (this work) & 81M & FBANK     & \xmark  & 7.2   \\

\hline

\hline
\end{tabular}
\\$^\dagger$ This model used speaker inventory as an auxiliary input.
\vskip-2mm
\end{table}

Table~\ref{tab:asr} shows the multi-talker speech recognition results on the {\tt test-clean} evaluation set with the SURT model shown in Fig.~\ref{fig:surt}-a). Previously, we achieved 10.8\% WER with Short-time Fourier Transform (STFT) features in~\cite{lu2020streaming}, and with FBANK features, we obtained slightly lower WER of 10.3\%. Our streaming results compares favorably with the offline sequence-to-sequence (S2S) based models~\cite{kanda2020joint, kanda2020minimum} and the RNN-T-based streaming model proposed in \cite{sklyar2020streaming}. 
We also evaluated the accuracy of an offline SURT model with bi-directional LSTMs (BLSTMs). For a fair comparison, we did increase the BLSTM model size by reducing the hidden size to be 600. The offline SURT model achieved 7.2\% WER, demonstrating the large gap between the offline and streaming SURT models. 

\begin{table}[t]\centering
\caption{Joint ASR and SID results. }
\label{tab:joint}
\footnotesize
\vskip-2mm
\begin{tabular}{cccc|cc}
\hline 

\hline
    System    & Feature & Training & SpecAug & WER  & SER \\ \hline
 B1    & FBANK &  Joint & \xmark & {\bf 10.1}  & {\bf 6.7}   \\
 B2          & FBANK & Stepwise &  \xmark &  10.3  &  8.7 \\
 B3           & FBANK &  Joint &  \cmark & 10.8  & 7.9   \\
 B4    & FBANK & Stepwise &  \cmark & 10.5  & 8.2   \\

\hline

\hline
\end{tabular}
\vskip-6mm
\end{table}

\subsection{Joint ASR and SID Results}

Table~\ref{tab:joint} shows the results of SURIT with or without joint training of the ASR and SID modules. For joint ASR and SID training, the SID loss is much smaller than ASR loss, and similarly for the $\ell_2$ norms of the gradients. To compensate for the difference, we set $\lambda=10$ and the results are given in Table~\ref{tab:joint}. We also evaluated the stepwise training approach, in which we trained the model with the ASR loss first, and then froze the Unmix module before training the model with SID loss only. The stepwise training approach is not as effective as the joint training approach in terms of accuracy as Table~\ref{tab:joint} shows. The reason may be that the Unmixing module was not trained with SID loss, and thus the features from the Unmixing module are not very discriminative for speakers. However, this approach has the advantage that it is suitable for training the model with transcription-only or speaker label only data. 
We also evaluated adding the SpecAugment~\cite{park2019specaugment} as an additional regularization approach on top of dropout and speed perturbation. For SID, this approach is effective for stepwise training, reducing the SER from 8.7\% to 8.2\%. However, for the joint training approach, additional regularization did not yield improvement, since multi-task learning itself is a kind of regularization. 

\subsection{SID Latency}

Finally, we evaluated the SID latency by measuring the average time step $t_e$ that HAT emitted the speaker label of the first utterance in an overlapped audio during inference. From the baseline system B4 in Table~\ref{tab:latency}, we observe that HAT delays the output of the speaker label until almost at the end of the audio segment. While this is not unexpected, practical applications usually favor low SID latency. To this end, we evaluated the two approaches discussed in Section~\ref{sec:latency}. To rule out the effect of the ASR module, we firstly used B4 in Table~\ref{tab:joint} as the baseline system, and tuned the two hyper-parameters $\alpha$ and $\beta$ while freezing the Unmixing module. From the results in Table~\ref{tab:latency}, we observed that a smaller $\alpha$ resulted in a significant latency reduction, albeit with the increase of SER. 

We then evaluated the latency penalty approach as Eq.~\eqref{eq:latency}. We set $t_{\text{buffer}} = 3$ for all our experiments. However, this approach itself did not work well in our experiments (C4 vs. B4), as this approach further reduces the gradient of speaker labels during training, and consequently it increases the probability to emit $\langle bs \rangle$ during inference. To mitigate this effect, we scaled down the gradient of $\langle bs \rangle$ with a smaller $\alpha$ during training, and we can achieve further latency reduction  (C5 vs. C1)). Finally, we performed joint training as Eq. \eqref{eq:joint_tr} with different values of $\alpha$ and $\beta$. Again, we set $\lambda = 10$ for these experiments. We observed a similar trend (D1-3 vs. B1), though the latency reductions were much more significant since the Unmixing module was also updated with the SID loss during joint training.





\begin{table}[t]\centering
\caption{Speaker identification latency. $t_e$ indicates the time step that emits the first speaker label, and $T$ is the length of the overlapped audio. }
\label{tab:latency}
\footnotesize
\vskip-2mm
\begin{tabular}{ccc|ccc}
\hline 

\hline
    System    & $\alpha$ & $\beta$ & SER    & $t_e$ & $t_e/T$ \\ \hline
  B4    & 1.0 & 0 & 8.2 &  181.7    & 0.95   \\
  C1    & 0.8 & 0 & 9.5 &  123.5    & 0.71   \\
  C2    & 0.6 & 0 & 9.9 &  103.7    & 0.62   \\
  C3  & 0.4 & 0 & 10.2 & 75.3 & 0.47 \\
  C4 & 1.0 & 1 & 8.5 & 173.1 & 0.92 \\ 
  C5 & 0.8 & 1 & 9.3 & 88.1 & 0.53 \\ 
  C6 & 0.8 & 3 & 10.6 & 62.4 & 0.37 \\ 
C7 & 0.8 & 5 & 10.1 & 96.9 & 0.58 \\ \hline
B1 & 1.0 & 0 & 6.7 & 153.8 & 0.85 \\
D1 & 0.8 & 0 & 7.1 & 55.4 & 0.34 \\
D2 & 0.6 & 0 & 7.6 & 52.7 & 0.32 \\
D3 & 0.8 & 1 & 7.9 & 54.8 & 0.33 \\
\hline

\hline
\end{tabular}
\vskip-6mm
\end{table}

\section{Conclusions}
\label{sec:conc}

In this paper, we aim at a simpler, end-to-end framework for rich transcription of meetings and conversational speech. In particular, we focused on joint ASR and SID from multi-talker audio in a streaming fashion. We proposed the SURIT framework, with the Umixing module as the feature extractor, while ASR and SID are both based on RNN-T and its variant. We proposed a transducer-based SID model, and evaluated two methods to reduce the SID latency. We studied the joint training approach for streaming ASR and SID, and demonstrated the effectiveness of the new framework with strong results. Our future work will focus on reducing the SID latency further as well as a large scale study of the proposed model.

\bibliographystyle{IEEEtran}

\bibliography{bibtex}

\end{document}